\documentstyle[prb,aps,multicol,epsfig,amssymb]{revtex}
  \renewcommand{\narrowtext}{\begin{multicols}{2} \global\columnwidth20.5pc}  
  \renewcommand{\widetext}{\end{multicols} \global\columnwidth42.5pc}  
  \newcommand{\wide}{\widetext \noindent \line(200,0){245} \line(0,1){3}\\}  
  \newcommand{\narrow}{\begin{flushright}\mbox{\line(0,-1){3}$\! \!$  
        \line(1,0){245}} \end{flushright} \narrowtext \noindent}

\newcommand{\pd}[2]{\frac{\partial #1}{\partial {#2}}}  
\renewcommand{\Re}{{\rm Re \,}}  
\renewcommand{\Im}{{\rm Im \,}}  

\begin{document}  
  
\title{Acoustoelectric current for composite fermions}  
\author{J. Bergli$^{(a)}$ and Y. M.  Galperin$^{(a,b,c)}$}  
\address{$^{(a)}$Department of Physics, University of Oslo, Box 1048 Blindern, 
 N-0316 Oslo, Norway;\\$^{(b)}$Centre for Advanced Studies, Drammensveien 78,  
0271 Oslo, 
Norway;\\$^{(c)}$Solid State Division, A. F. Ioffe   
Physico-Technical  Institute, 194021 St. Petersburg, Russia}  
\maketitle  
\begin{abstract}   
The acoustelectric current for composite fermions in a two-dimensional
electron gas (2DEG)  close   
to the half-filled Landau level is calculated in the random phase
approximation. The  
Boltzmann equation is used to find    
the nonequilibrium distribution of composite fermions to second order
in the acoustic field.    
It is shown that  the oscillating Chern-Simons field created by the   
induced density fluctuations in the 2DEG 
is important for the acoustoelectric  
current. This leads to a violation of the Weinreich 
relation between the acoustoelectric current and acoustic intensity.
The deviations from the  Weinreich relation can be    
detected by measuring the angle between the longitudinal and the Hall
components of the acoustoelectric current. This departure from the Weinreich   
relation gives additional information on the properties of the 
compostite fermion fluid.
\end{abstract}  
  
  
\narrowtext  
\section{Introduction}  
  
Two-dimensional electron gases (2DEG) have been studied extensively both   
experimentally and theoretically. One important experimental   
technique is to investigate the interaction of the electron gas   
with surface acoustic waves (SAW) propagating along the sample. Because of   
the piezoelectric properties of the substrate materials (GaAs-AlGaAs),   
the acoustic wave is accompanied by an electric wave that interacts with   
the electron gas. A traveling wave of electric field can also be 
produced by placing a 2DEG sample on the surface of a piezoelectric crystal. 
  
It is well known  
that both the attenuation and sound velocity are sensitive to changes in the   
properties of the electron   
gas.~\cite{Wixforth1986,Wixforth1989,Willett1993,Willett1995,Halperin1993,Simon1996,Knabchen1996,Drichko1999}
 Less well explored experimentally   
is the acoustoelectric drag current induced by the acoustic wave.   
{}From the experiments so far, it seems that in some cases   
this offers the promise of greater sensitivity than the attenuation or  
velocity shift measurements, because one measures directly a small  
electric current or voltage  
instead of a small shift in a large quantity.\cite{Shilton1995b}    
In addition, measurements of the acoustoelectric current gives a   
direct measure of the intensity of the electric field created by the   
acoustic wave, thus enabling determination of the coupling between the   
SAW and the electron gas, whereas attenuation measurements only give   
relative intensities.\cite{Wixforth1986,Wixforth1989}   
The theoretical efforts in this direction have been modest   
\cite{Efros1990,Falko1993} and a better understanding is required to   
interpret measurements.\cite{Shilton1995a} Furthermore, there is some   
disagreement between the different methods,\cite{Efros1990,Falko1993} that   
has not been clarified yet.   
  
The properties of a 2DEG in a strong magnetic field have successfully   
been described by the composite fermion model.\cite{Halperin1993,Mirlin1997b}  
Especially near even-denominator filling fractions, like $\nu_f=1/2$, this   
seems to be a good description. In this paper,  
we will calculate the acoustoelectric   
current for composite fermions using the Boltzmann equation approach.  
This has previously been applied to the calculation of the conductivity   
tensor at finite wave vector and frequency.\cite{Mirlin1997b}   
The paper is organized as follows. In Sec.~\ref{int} an interaction  
between composite fermions and a SAW will be discussed. The Boltzmann  
equation is derived and solved in Sec. \ref{accur}, and the   
acoustelectric current is calculated. The resulting expression   
is discussed in Sec. \ref{discussion}.

\section{Interaction between composite fermions and a surface  
acoustic wave} \label{int}  
  
There are several fields involved in the problem.\cite{Simon1998}   
The real physical fields, those seen by the electrons, are   
the external magnetic field $\bf B$, and a periodic electric   
field set up by the acoustic wave.  
  
The materials used to create the 2DEG, GaAs-AlGaAs heterostructures, are   
piezoelectric. This means that an acoustic wave propagating on the sample   
will create an periodic electric wave which interacts with the composite   
fermions. To achieve stronger coupling, one sometimes uses a substrate   
with a larger piezoelectric constant, and places the heterostructure in   
close contact with (but acoustically isolated from) the substrate. The   
piezoelectric field induced by the SAW is then able to penetrate   
the 2DEG.\cite{Wixforth1989}   
  
It is assumed that the wave is propagating in the $x$-direction,   
and that the piezoelectric field is in the same direction. The   
electric field is then given by the real part of   
${\bf E} ({\bf r}, t) = {\bf E}_0 e^{-i(\Omega t -   
{\bf qr})} = -\nabla \Phi$ with   
$\Phi = \Phi_0  e^{-i(\Omega t -   
{\bf qr})}$. Here $\Omega,{\bf q}$ are the SAW frequency and wave  
vector, respectively; ${\bf E}_0 \parallel{\bf q} \parallel {\bf  
\hat{x}}$. $\Phi_0$ is the amplitude of the screened potential, it is   
related to the amplitude of the   
piezoelectric (external) potential $\Phi_{\text{ext}}$ by the   
equation (see Ref. \onlinecite{Simon1996}):   
$\Phi_0=[1-v(q)K_{00}]\Phi_{\text{ext}}$, where
$v(q)=2\pi/\epsilon_{\text{eff}} q $ is the Fourier  
transform of    
the Coulomb potential and the response function is   
 \begin{equation}\label{s} 
 K_{00}(q,\Omega)=q \frac{\sigma_m}{v_s} \frac{1}{(1-i\sigma_m/\sigma_{xx}^e 
)}    
 \end{equation}  
where $\sigma_{xx}^e({\bf q},\Omega)$ is the longitudinal electron 
conductivity,  $\sigma_m \equiv v_s\epsilon_{\text{eff}}/2\pi$, $v_s =
\Omega/q$
is the SAW velocity, and 
$\epsilon_{\text{eff}}$ is the effective    
dielectric constant. The amplitude of the piezoelectric field is related   
to the amplitude, $A$, of the acoustic wave by the relation   
 \begin{equation}  
 \Phi_{ext}=Ae_{14}F(qd)/\epsilon,  
\end{equation}  
where $e_{14}$ is the piezoelectric stress constant and $F(qd)$ is some  
dimensionless function calculated by Simon.\cite{Simon1996}  
  
When we make the Chern-Simons (CS) transformation, additional,
fictitious, fields are introduced. There will be a Chern-Simons
magnetic field ${\bf b}= -2\phi_0\,n_e^{\text{tot}}{\bf \hat{z}}$,
where $\phi_0= 2\pi\hbar c/e$ is the flux quantum and
$n_e^{\text{tot}}$ is the total electron density. Because of the
interaction with the piezoelectric field, there will be an induced
density modulation in the electron gas. Therefore, the electron
density is conveniently split in two parts, an average and a
fluctuating (AC) part: $n_e^{\text{tot}}=n_e+\delta
n_e$. Corresponding to these, there will be an average and an AC
Chern-Simons magnetic field. The average field will partly cancel the
external (real) magnetic field, leaving the effective field ${\bf
B}_0^*={\bf B}(1-2\nu_f)$, where $\nu_f=\phi_0n_e/B$ is the Landau
level filling factor.  For simplicity we assume two flux quanta to be
attached to each electron, appropriate for $\nu_f$ close to $1/2$.
The AC component of the Chern-Simons field given by ${\bf
b}^{\text{ac}}= -2\phi_0\,\delta n_e {\bf \hat{z}}$.  In addition, the
motion of the CFs will create an electric Chern-Simons field which is
given by ${\bf e} = \rho\,({\bf \hat{z}}\times{\bf j})$, where
$\rho=2\phi_0/ec=2h/e^2$.  The current is split 
in two parts, an AC part ${\bf j}^{\text{ac}}$,
which is given by the linear response, and a DC part, the acoustoelectric 
current ${\bf j}^{\text{ae}}$, which is given by the second order response. 
These give contributions ${\bf e}^{\text{ac}}$ and ${\bf e}^{\text{ae}}$ 
to the CS electric field.
  
The strength of these fields relative to the perturbing field $E =  
|{\bf E}|$ can    
be estimated from charge conservation. The equation for conservation 
of charge    
is $e  \,\partial(\delta n_e)/\partial t + \nabla {\bf j}=0$, or Fourier transformed,  
$ -i\Omega e \delta n_e + i \sigma_{xx}^e q E_x = 0 $. This gives  
\begin{equation}  
 -e\,\delta n_e = (\sigma_{xx}^e/v_s)\,E_x\, .  
\end{equation}   
Then the force from the fluctuating (AC) component of the magnetic  
Chern-Simons field is given by  
\begin{equation}\label{estimate}  
 \frac{ev}{c}\,b^{\text{ac}} = -2\phi_0 
  \frac{v}{c}\frac{\sigma_{xx}^e}{v_s}E_x    
  =\frac{4\pi}{\alpha}\frac{v}{c}\frac{\sigma_{xx}^e}{v_s}\,eE_x,  
\end{equation}  
where $\alpha=e^2/\hbar c=1/137$ is the fine structure constant.  
In order for the   
acoustic wave to be sensitive to changes in the properties of   
the electron gas, we see from Eq.~(\ref{s}) that  
the ratio of the conductivity to the  
sound velocity should be $|\sigma_{xx}^e|/v_s\approx 
\epsilon_{\text{eff}}/(2\pi)$  
(or at least not differing   
by more than one order of magnitude, see  
Refs.~\onlinecite{Wixforth1986,Shilton1995b,Shilton1995a}).  
With $\epsilon_{\text{eff}}\approx 6$  
(see Ref. \onlinecite{Simon1996}), we have 
$|\sigma_{xx}^e|/v_s\approx 1$. It is then  
seen that the relative strength of the force from the AC Chern-Simons   
field and the direct force from the piezoelectric field of the SAW is    
$(4\pi/\alpha)\, (v/c)$. With a Fermi velocity of $10^5$ m/s, this   
factor is of order 1. That is, the piezoelectric  and the CS fields are   
of comparable importance.  
For the electric CS field we find for the $x$-component 
\begin{equation} 
 e_x^{\text{ac}}=- \rho \sigma_{xy}^e ({\bf q}, \Omega) E_x \,. 
\end{equation} 
Since static $\sigma_{xy}^e \sim e^2/2h$ for 
$\nu_f\approx 1/2$,   
we see that the Chern Simons field is of the same order as the  
external electric field.
The $y$-component is similarly  
\begin{equation}  
 e_y^{\text{ac}}= \rho |\sigma_{xx}^e| E_x 
   \approx \rho v_s E_x  
  =(4\pi/\alpha)\, (v_s/c)\, E_x  
\end{equation}  
which is smaller than what is given by equation (\ref{estimate})  
by the factor $v_s/v$.
Since the typical sound velocity   
is $3\times 10^3$ m/s, this will be small. However, it is acting 
along the trajectory of the composite fermions at the points of 
strong interaction (see below), and will therefore play an important 
role.  

So far we have expressed the current as response  
to the external field.  
However, as seen from the composite fermions it is not  
possible to separate the two fields, and it is more convenient to consider  
the response to the total effective electric field acting on the  
composite fermions, ${\bbox {\cal E}}=\bbox{E+e}^{ac}$. The latter is 
described by the   
composite fermion conductivity tensor, $\sigma_{ik}({\bf q},\Omega)$.  
Because the field is not 
parallel to the wave vector, this is not a potential field, and we can 
not write it as the gradient of a potential. However, we can consider the 
$x$-component $ {\cal E}_x=E_x+e_x^{\text{ac}} =
 -\nabla_x
\Psi$.
In the following we will consider the response to the effective field 
$\bbox{\cal E}_x$ acting upon composite fermions. 
The only thing that needs to be changed in order to account for  
the Chern-Simons field is then that the induced density modulation  
must be expressed in terms of response to the total field,  
\begin{equation}  
 e\delta n_e=v_s^{-1}\left(\sigma_{xx}{\cal E}_x+\sigma_{xy}{\cal E}_y 
   \right). 
\end{equation} 
The $y$-component of the total field is just the Chern-Simons field  
since $E_y=0$, and one has 
\begin{equation} 
 {\cal E}_y=\rho\left(\sigma_{xx}{\cal E}_x+\sigma_{xy}{\cal 
 E}_y\right)\, . 
\end{equation} 
 {}From this we get  
\begin{equation} 
 {\cal E}_y=\delta{\cal E}_x\, , 
\end{equation} 
where  
\begin{equation} 
 \delta=\rho\tilde{\sigma}e^{i\eta}=\rho\sigma_{xx}/(1-\rho\sigma_{xy})\,
 .
\label{delta1}
\end{equation}
We then have 
\begin{equation} 
 b^{ac}=- (c/v_s)\, \delta 
   e{\cal E}_x \, .
\end{equation} 
In the most interesting situation 
the acoustic wave length $2\pi/q$ is less than the typical diameter of 
CF orbits. That is, the parameter $\kappa=qR_c$ is large, where $R_c$  
is the radius of the cyclotron orbit. Using the expression for  
$\sigma_{xy}$ obtained by Mirlin and W\"{o}lfle \cite{Mirlin1997b} 
we find in this limit 
\begin{equation}\label{sigmaxy} 
 \rho\sigma_{xy}\simeq\frac{1}{\zeta}\frac{\cos 
   2\kappa}{\sin\pi(\omega+i\nu)}\, , \qquad
   \zeta=\frac{\hbar\Omega}{2mv_s^2}\, . 
\end{equation} 
Here we have introduced dimensionless frequency  
$\omega=\Omega/\omega_c$ and damping $\nu=(\omega_c \tau)^{-1}$, 
 where  
$\omega_c \propto B^*$ is the CF cyclotron frequency. 
Substituting reasonable values,  
$\Omega/2\pi = 3$ GHz, $v_s=3\times 10^3$ m/s, $m \approx 
10^{-27}$ g and $\nu\approx 1$, we obtain a quantity of the order 1  
out of cyclotron resonance, and an enhancement of the order  
$\kappa$ at cyclotron resonance, where $\omega=n$ is an integer. 
This means that the factor $(1-\rho\sigma_{xy})^{-1}$
may be arbitrary.   
 
In the following we shall use the random phase approximation (RPA), 
according to which the composite fermions are considered as 
non-interacting particles. It is known \cite{Simon1998,Stern1995} that if 
we are to have a consistent Fermi liquid theory we must also include the Landau 
interaction parameters. This will affect the CF mass, and and will possibly 
give some additional effects. This is an interesting possibility for  
further work.~\cite{end1}  
 
Even within the RPA, 
the problem is not equivalent to the  
corresponding    
problem for electrons in the effective magnetic field ${\bf B_0^*}$, because  
of the fluctuating Chern-Simons fields associated with the   
density modulation induced by the SAW. 
 
\section{Acoustoelectric current} \label{accur}  
\subsection{Solution of the Boltzmann equation} \label{solbe}  
The acoustoelectric current will be expressed through the  
nonequilibrium distribution function for composite fermions, $f({\bf  
r},{\bf p},t)$. In this   
section, we will calculate this function from   
the composite fermion Boltzmann equation.   
As long as the amplitude of the acoustic wave is sufficiently small   
we can treat the induced piezoelectric field as a perturbation. That means   
that the Boltzmann equation can be first linearized and then solved  
iteratively. Thus, we seek a solution to the equation  
\begin{equation}  
 \left(\partial/\partial t + {\bf v}\nabla_{\bf r} +   
  {\bf F}\nabla_{\bf p}\right) f({\bf r},{\bf p},t) = C\{f\}  
\end{equation}  
of the form   
\begin{eqnarray} \label{expansion}  
f&=&f_0[\epsilon_{\bf p}+e\Psi ({\bf r},t)]+ e\Psi\left( -  
\partial f_0/\partial\epsilon\right)f_1({\bf r},t) \nonumber \\  
&& \quad \quad-  
\left(e^2|\Psi_0|^2q/2m\right)f_2({\bf r},t)\, ,   
\end{eqnarray}  
where $f_0$ is the unperturbed solution (with ${\bf E} = 0$ but the   
constant magnetic field included).   
 The force must be taken as  
\begin{equation}  
  {\bf F}=-e\bbox{\cal E}-(e/c)\left[{\bf v}\times({\bf B}_0^* + {\bf 
  b}^{\text{ac}})\right] \, . 
\end{equation}  
  The important scattering mechanism for composite fermions is scattering   
by the random magnetic (Chern-Simons) field which is created by density   
fluctuations in the electron gas because of electrostatic potentials from   
impurities in the doping layer.\cite{Halperin1993} Further, it is known   
that to ensure particle number conservation, care has to be taken in writing  
the collision operator, as emphasized by Mirlin and  
W\"{o}lfle.\cite{Mirlin1997b}   
However, the details of the scattering mechanism are    
not expected to change the results in a qualitative way   
(see, e. g., Ref.~\onlinecite{Simon1998}), and for simplicity we will  
use the relaxation time  approximation $C\{f_1\}=-f_1/\tau$. This will  
greatly simplify the calculations.    
 
The Boltzmann equation is most conveniently expressed in terms of polar   
coordinates ($v$,$\phi$) in the ($v_x$,$v_y$) plane, where $v=|{\bf v}|$   
is the absolute value of the velocity and $\phi$ is the angle between   
the velocity and the $x$-axis.   
In these coordinates the linearized equation has the form   
$ {\hat L}_\omega f_1 = i \omega+i\kappa\delta\sin\phi$,  
where   
\begin{equation}\label{B1}  
 {\hat L}_\omega=\partial/\partial\phi+\nu + i(\kappa\cos \phi -\omega)\, .  
\end{equation}  
 The quantity  
$2 \pi \kappa$ has a meaning of the ratio of the CF cyclotron  
radius to the acoustic wavelength.

The linearized equation has the periodic solution  
\begin{eqnarray}\label{sle}  
f_1=&& \frac{i\omega}  {e^{2\pi (\nu - i \omega)}-1}
   \int_\phi^{\phi +2\pi}\! \!d\phi' \, \nonumber\\
 &&(1+\beta\delta\sin\phi')  
  e^{(\nu - i\omega)(\phi'-\phi)+i\kappa\, (\sin  
   \phi'-\sin\phi)}\, .   
\end{eqnarray}  
The second order approximation  will have components with frequency 0 and   
$2\Omega$. Since we are only interested in the DC acoustoelectric current,   
it is sufficient to extract non-oscillating part of the response.   
Thus for the second iteration the equation can be simplified to  
\begin{equation}  
 {\bf F}\nabla_{\bf p} f({\bf r},{\bf p},t) = C\{f\}.  
\end{equation}  
In addition, since we are using complex notation, to get the DC component  
from the term $\propto \bbox{\cal E}\nabla_{\bf p}f_1$ we have to take the  
complex conjugate of the electric field and divide by 2. That is, the   
non-oscillatory component of the product of two oscillating functions,  
$A({\bf r},t) \equiv \Re \left[A_1 \, e^{-i(\Omega t - {\bf qr})}\right]$ and   
$B({\bf r},t) \equiv \Re \left[B_1 \, e^{-i(\Omega t - {\bf qr})}\right]$   
is equal to $\Re (A_1B_1^*/2)$.  
  
Writing the second order approximation according to Eq.~(\ref{expansion})  
we get the linear equation
\begin{equation}\label{sob}
 {\hat L}_0f_2=S(v,\phi) - \frac{2m\rho}
 {e|\Psi_0|^2q\omega_c}\pd{f_0}{\epsilon}\,
  {\bf v}\cdot(\hat{\bf z}\times{\bf j}^{\text{ae}})\, .
\end{equation}
 Here 
${\hat L}_0 =\partial/\partial \phi+\nu$,  
\begin{eqnarray}   
 S&&(v,\phi)=-\frac{1}{\omega_c}\left[\left(\cos \phi \pd{}{v}  
 -\frac{1}{v}\sin \phi   
 \pd{}{\phi}\right)\left(\pd{f_0}{\epsilon}  
 \,\Im f_1 \right)\right. \nonumber \\   
&& \left. +\rho\tilde{\sigma}\left[\sin \phi \pd{}{v}  
  +\frac{1}{v}\cos \phi \pd{}{\phi}\right]\left(\pd{f_0}{\epsilon}  
  \Im\{e^{-i\eta} f_1 \}\right)\right. \nonumber\\
&&  \left.- \frac{\beta\rho\tilde{\sigma}}{v} 
 \pd{}{\phi}\, \left( \pd{f_0}{\epsilon}\,  
 \Im\{e^{i\eta} f_1\} \right) \right]\, , \label{sv}  
\end{eqnarray}  
where $\beta=v/v_s \gg 1$.  
The last term in the $S(v,\phi)$ are the CS-contributions.   
The periodic solution for $f_2$ is  
\begin{eqnarray}\label{sbs}
 f_2(v,\phi)&=&\frac{1}{e^{2\pi\nu}-1}\int_\phi^{\phi +2\pi}\! d\phi'\,
  S(v,\phi') \,e^{\nu( \phi'- \phi)} \nonumber\\
  &&-  \frac{2m\rho}
  {e|\Psi_0|^2q\omega_c}\pd{f_0}{\epsilon}\,
  \frac{1}{e^{2\pi\nu}-1}\nonumber\\
  &&\quad  \times\int_\phi^{\phi +2\pi}\! d\phi'\, {\bf
  v}\cdot(\hat{\bf z}\times{\bf j}^{\text{ae}})e^{\nu( \phi'- \phi)}
  \, .
\end{eqnarray}  
This is the solution of the Boltzmann equation to second   
order in the perturbation, from which we now will calculate   
the acoustoelectric current.   
  
\subsection{Calculation of the DC current}  
  
Neither the equilibrium distribution nor the first order   
perturbation, $f_1$, will give any contribution to the   
DC current, so the lowest order contribution is found from   
the second order perturbation. Let us for the moment forget the 
last term in the expression (\ref{sbs}), which comes from the 
Chern-Simons field created by the DC acoustoelectric current, 
and calculate the current $   {\bf j}_0^{\text{ae}}$ from the 
first term. 
 \begin{eqnarray}\label{j}  
 \left\{\begin{array}{c}j_{0,x}^{\text{ae}}
  \\j_{0,y}^{\text{ae}}\end{array}\right\}&=&  
\frac{e^3|\Psi_0|^2q}{2m}\int_0^{\infty} \! dv\,  
    v^2 \int_0^{2\pi}d\phi  
    \left\{\begin{array}{c}\cos \phi \\ \sin \phi  \end{array} \right\}  
\nonumber \\ && \times    
  \int_{\phi}^{\phi+2\pi}\ d\phi' \, \frac{S(v,\phi')\, e^{\nu (\phi'-  
\phi)}}{(e^{2\pi\nu}-1)}\, .   
\end{eqnarray}  
The first term of $S(v,\phi')$, Eq.~(\ref{sv}), can be integrated by  
parts over $v$ to get    
\begin{equation} \label{cur1}  
\frac{e^3|\Psi_0|^2q}{2m\omega_c}\int_0^{\infty}  
    \!dv\,v\pd{f_0}{\epsilon}\int_0^{2\pi}d\phi   
    \left\{\begin{array}{c}\cos \phi \\ \sin \phi  \end{array}  
    \right\} \frac{F(\phi)}{e^{2\pi\nu}-1}  
\end{equation}  
where  
\begin{eqnarray}    \label{current}  
&&F(\phi) \equiv \int_{\phi}^{\phi+2\pi} \! d\phi' \, \left[2\cos\phi'\,  
    \Im f_1 + \sin\phi' \pd{\,\Im f_1}{\phi'}\right.  
\nonumber \\ && \qquad\left. + 2\rho\tilde{\sigma}\sin\phi'\,  
    \Im\{e^{-i\eta} f_1\}\right.\\ 
 &&\left. - \rho\tilde{\sigma}\cos\phi' 
    \pd{\,\Im \{e^{-i\eta} f_1\}}{\phi'}     
 +\beta\rho\tilde{\sigma}\pd{\,\Im\{e^{i\eta}  
    f_1\}}{\phi'}  
    \right]e^{\nu (\phi'- \phi)}.\nonumber
\end{eqnarray}  
Since $\Im f_1$ and $\Im\{e^{i\eta}f_1\}$ are periodic functions   
of $\phi$ they can be expanded in  Fourier series,  
$$
 \Im f_1=\sum_{n=-\infty}^{\infty}A_ne^{in\phi} \, , \quad   
 \Im \{e^{i\eta} f_1\}=\sum_{n=-\infty}^{\infty}\tilde{A}_ne^{in\phi}
\, ,   
$$
where the $A_n$ and $\tilde{A}_n$ will be functions of $v$.  
The angular integrals are then reduced to the six cases   
$$
 \int_0^{2\pi}d\phi  
   \left\{\begin{array}{c}\cos \phi \\ \sin \phi  \end{array} \right\}  
   e^{-\nu\phi}\int_{\phi}^{\phi+2\pi}d\phi'  
   \left\{\begin{array}{c}\cos \phi' \\ \sin \phi' \\ 1 \end{array} \right\}  
   e^{(\nu+i n)\phi'},  
$$
where all combinations of the expressions  
in the braces are implied. These can be evaluated directly, and   
it is found that only the zeroth, $A_0=\overline{\Im f_1}$, and first Fourier 
components will give contributions.   
The current is then given by  
\begin{equation} 
{\bf j}_0^{\text{ae}} 
=\frac{\pi  
e^3|\Psi_0|^2q}{m\omega_c(\nu^2+1)}\int_0^{\infty} \! dv\,v \,  
\pd{f_0}{\epsilon} \sum_{i=1}^4 {\bf J}_i (v) 
\end{equation}  
where  ${\bf j}_0^{\text{ae}} \equiv
 \left\{\begin{array}{c}j_{0,x}^{\text{ae}}\\j_{0,y}^{\text{ae}}\end{array}\right\}
 $, 
\begin{eqnarray}  
{\bf J}_1&=& A_0\left\{\begin{array}{c}\nu\\1  
\end{array}\right\} \, , \\  
{\bf J}_2&=&-\beta\rho\tilde{\sigma}\,  \Im \tilde{A}_1
 \left\{\begin{array}{c}\nu\\1\end{array}\right\}\,, \\
 {\bf J}_3&=&\rho\tilde{\sigma}\tilde{A}_0\left\{\begin{array}{c}-1\\\nu  
\end{array}\right\} \, , \\  
{\bf J}_4&=&-\beta\rho\tilde{\sigma} \, \Re \tilde{A}_1
 \left\{\begin{array}{c}-1\\\nu\end{array}\right\}\,.
\end{eqnarray}  
At low temperatures we may assume $\partial f_0/\partial\epsilon=-(m/  
4\pi \hbar^2) \,\delta(\epsilon-\epsilon_F)$,  
and we can perform the integral over $v$, which will   
fix $v$ at $v_F$ (and $\beta$ at $v_F/v_s$).   
The final result is then 
\begin{equation}\label{aecurrent}  
{\bf j}_0^{\text{ae}}=\frac{j_0}{\nu^2+1} \sum_{i=1}^4 {\bf J}_i (v_F)
\, , \quad j_0=\frac{e|\Psi_0|^2q}{\hbar\omega_c\rho}\, .
\end{equation}  
We must now return to the last item in Eq.~(\ref{sbs}). Since ${\bf
j}^{\text{ae}}$ is a constant, the integral may be evaluated directly,
and the resulting equation solved for ${\bf j}^{\text{ae}}$ in terms
of ${\bf j}_0^{\text{ae}}$.  There is, however, a simpler and
physically more transparent way of obtaining the same result.  We
calculate ${\bf j}^{\text{ae}}_0$ as before. Then we write ${\bf
j}^{\text{ae}}={\bf j}^{\text{ae}}_0+\delta{\bf j}^{\text{ae}}_0$, where 
$\delta{\bf j}^{\text{ae}}$ is the response to the CS electric field 
created by the acoustoelectric current,
\[
 \delta{\bf j}^{\text{ae}}=\hat{\sigma}{\bf
  e}^{\text{ae}}=-\sigma\hat{\rho}_{CS} 
  {\bf j}^{\text{ae}}, 
\]
where 
\[
\hat{ \rho}_{CS}=\rho\left(\begin{array}{cc} 0 &1\\-1&0\end{array}\right).
\]
Using the Drude conductivity for $\sigma$ we have 
\begin{equation}\label{DCcond}
\hat{ \sigma}=\frac{\sigma_0\nu}{1+\nu^2}\left(\begin{array}{cc} 
   \nu &-1\\1&\nu\end{array}\right), \qquad\qquad \sigma_0=\frac{n_e 
  e^2\tau}{m}\, ,
\end{equation}
which gives 
\begin{equation}\label{ren}
 {\bf j}^{\text{ae}}=(1+\hat{M})^{-1}{\bf j}^{\text{ae}}_0, 
\end{equation}
where 
\[
\hat{ M}=\hat{\sigma} \hat{\rho}_{CS}=\frac{\gamma}{1+\nu^2}\left(\begin{array}{cc} 1
 &\nu\\-\nu&1\end{array}\right), \qquad
 \gamma=\frac{2\epsilon_F}{\hbar\omega_c}=2\nu_f^*.
\]
Here $\nu_f^*$ is the filling factor of composite fermions (the filling 
factor in the effective field $B_0^*$).

\subsection{Results}  
The Fourier coefficients $A_n$   
\begin{equation} \label{a_n}  
 A_n=\frac{1}{2\pi}\int_0^{2\pi}\! d\phi\,  e^{-in\phi}\, \Im f_1 \, ,  
\end{equation}  
and  $\tilde{A}_n$, given by a similar expression,  cannot be evaluated  
in closed form for arbitrary $\omega$ and $\kappa$. However, in the  
most interesting situation the acoustic wavelength appears much   
smaller 
 than the CF cyclotron radius, and $\kappa \gg 1$. In this  
paper we restrict ourselves to this case.  
It will be shown later that our calculation method remains valid  
provided $\nu \lesssim (1/\pi) \ln \kappa$.   
Then the coefficients $A_n$ and  $\tilde{A}_n$  can be evaluated  
using the method of stationary phase. The critical points are   
approximately  
$\phi=\pm \pi/2$, which  is consistent with the physical   
picture that the CF will interact significantly with the acoustic wave only   
when the electron momentum is normal to the direction of propagation of the   
wave. For other directions of the momentum, the CF will be subject to   
a rapidly oscillating force giving no net contribution (see
Fig.~\ref{cf1}).
\input{fig1.inp}  
 It is not convenient to calculate these coefficients directly,   
because the expression for $\Im f_1$ is complicated. A simpler way   
is to calculate the coefficients   
$$
 B_n=\frac{1}{2\pi}\int_0^{2\pi} \!d\phi\, e^{-in\phi}\, f_1 \, ,  
$$
and then calculate $A_n$ as  $(B_n-B_{-n}^*)/2i$, and similarly 
for $\tilde{A}_n$.
The result is (see appendix \ref{asymptotic})  
\begin{eqnarray}
A_0&=&\Re\left[\frac{ 1+z^2 
  +2z (\sin 2\kappa-i\beta\delta\cos 2\kappa)}
  {\beta(z^2-1)}  
 \right]\, , \label{A0}\\ 
\tilde{A}_0&=&
    \cos\eta\, \Re B_0+\sin\eta\, \Im B_0\, , \\
\tilde{A}_1&=& -\frac{2i}{\beta}\cos2\kappa \,
  \Im\left[\frac{ze^{-i \eta}}{z^2-1}\right] -
  \frac{1}{\beta}\sin\eta \nonumber \\
& -&i\rho\tilde{\sigma}\, \Re\left[\frac{z^2+1}
   {z^2-1}\right] + 2i\rho\tilde{\sigma}\sin 2\kappa\,
  \Re\left[\frac{z^2} 
   {z^2-1}\right] \, ,\label{A1}
\end{eqnarray}
where $z=e^{\pi(\nu -i \omega)}$.
These expressions are then to be inserted into the formula (\ref{aecurrent})
for the acoustoelectric current. 

In the last expressions we have neglected terms suppressed by factors of   
$\beta^{-1}$ or $(\beta\omega)^{-1}=\kappa^{-1}$, even if these where  
of larger power in  $e^{\pi\nu}$. This is justified as long as  
$e^{\pi\nu}$ is smaller than   
$\kappa$, or $\nu \lesssim (1/\pi)\ln \kappa$. This inequality sets  
the limit of applicability of the stationary phase approximation.

\section{Discussion}\label{discussion}

The final expressions (\ref{A0})-(\ref{A1}) are not so simple, 
and we will try to understand how they behave as the external 
magnetic field is changed. Also, we will see how this affects the 
acoustoelectric current (\ref{aecurrent}). 

The Fourier coefficients show two kinds of oscillations: geometric
oscillations and cyclotron resonance. 
We analyze the behavior of these as functions of  
increasing dimensionless parameter  
$\omega$. This corresponds to decreasing effective magnetic field at  
fixed acoustic frequency $\Omega$.  
As the magnetic field changes, the value of  
$\nu=(\tau\omega_c)^{-1}=\tilde{\nu}   
\omega$ will also change. Here $\tilde{\nu}=(\Omega\tau)^{-1}$.  
Consider for example $A_0$, Eq.~(\ref{A0}). First, there are the terms $\sin 2\kappa$ and $\cos
2\kappa$, which gives geometric oscillations as for metals.
\cite{Bommel1955,Pippard1957,Gurevich1960} The oscillations arise as
the difference in the phase of the sound wave at the two points of the
cyclotron orbit where the CF's interact efficiently with the sound
wave is changing. There is a complete oscillation as the diameter of
the cyclotron orbit increases by one wavelength ($2\kappa$ increases
by $2\pi$).  On top of this comes oscillations from the
$e^{i\pi\omega}$ terms.  Since $\beta\approx 30\gg 1$, these
oscillations are much slower. They describe cyclotron resonances,
which are determined by the relative phase of the sound wave as the CF
pass through the same point of the cyclotron orbit at successive
revolutions. If the CF experiences the same phase every time it passes
a specific point, it will resonate, and the interaction will be
strong. This happens when the acoustic frequency is an integer
multiple of the cyclotron frequency, i.e. when
$\omega=\Omega/\omega_c=n$ is an integer. Finally there will be an
overall damping as $\nu$ is growing, so that for $e^{\pi\nu}\gg 1$ no
interesting behavior is expected.

This behavior is seen in
Fig.~\ref{A0graph}, which shows the expression 
\[ \cos 2 \kappa\,  \Re \frac{z }{z^2-1}=
 \cos 2 \kappa \, \Re\left[\frac{e^{\pi(\nu-i\omega)}}{e^{2\pi(\nu-i\omega)}-1}
  \right]
\]
which comes from the last term in the numerator of $A_0$, as a function
of $\omega$ for the case $\beta=30$ and $\tilde{\nu}=0.5$.  Our
approximations are valid for $\nu\lesssim(1/\pi)\ln\kappa$ and
$\kappa\gg 1$. For these values of the parameters, this corresponds to
the range $0.04<\omega<2.8$.
\input{fig2.inp}
We will now concentrate on the most realistic limit of large damping, 
$\nu>1$ (remember however that we must have $\nu<(1/\pi)\ln\kappa$ for 
our stationary phase approximation to be valid).
In this limit we have $1\ll e^{\pi\nu}\ll e^{2\pi\nu}$, and the 
Fourier components can be expanded in powers of $|z|^{-1}=e^{-\pi\nu}$. 
To lowest order we get 
\begin{eqnarray*}
 A_0&=&\beta^{-1} + 2\beta^{-1}e^{-\pi\nu}\\&&\times \left[\cos\pi\omega
  \cos2\kappa+\beta|\delta|\cos2\kappa\sin(\pi\omega+\eta)\right]\, , \\
\tilde{A}_0&=&\beta^{-1}\cos\eta 
   + 2\beta^{-1}e^{-\pi\nu} \\ && \times
\left[\cos(\pi\omega-\eta)\sin2\kappa 
   + \beta|\delta|\cos2\kappa\sin\pi\omega\right]\, ,\\
 \tilde{A}_1&=&-\beta^{-1} \sin\eta-i\rho\tilde{\sigma} 
  +2ie^{-\pi\nu}\\ && \times
\left[|\delta|\sin2\kappa\cos\pi\omega- \beta^{-1}
  \cos2\kappa\sin(\pi\omega-\eta)\right].
\end{eqnarray*}
We observe that the factor $\rho\tilde{\sigma}$ occurs repeatedly in these 
expressions, and also in the expression (\ref{aecurrent}) for the 
acoustoelectric current. We will therefore analyze this expression. 
To this end we recall the definition (\ref{delta1}) of the quantity $\delta$ 
and use the expressions obtained by Mirlin and W\"olfle for
$\sigma_{xx}$ and $\sigma_{xy}$, equation (8) of
Ref.~\onlinecite{Mirlin1997b}. In the limit $\kappa\gg1$ we can expand the 
Bessel functions in asymptotic series in $1/\kappa$ and obtain
$ \rho\sigma_{xx}=-i/\zeta$, 
and $\rho\sigma_{xy}$ is given by Eq.~(\ref{sigmaxy}).
This gives
\begin{equation}\label{delta}
 \delta=-i\left[\zeta-\frac{\cos2\kappa}{\sin\pi(\omega+i\nu)}\right]^{-1}
 \, .
\end{equation}

Substituting reasonable values,  
$\omega/2\pi = 3$ GHz, $v_s=3\times 10^5$ cm/s, $m \approx 
10^{-27}$ g we obtain $\zeta\approx1/10$. 
We consider now two cases 

\begin{enumerate}

\item {\em At cyclotron resonance}, where $\omega=n$ for an integer $n$. 
In this case we have $\sin\pi(\omega+i\nu)=\pm i\sinh\pi\nu$, the sign 
being $+$ or $-$ when $n$ is even or odd, respectively. In the denominator 
of Eq.~(\ref{delta}) we then have a real contribution from $\zeta$ and a 
purely imaginary contribution, $\pm i \cos2\kappa/\sinh\pi\nu$. 
Thus, we get 
\[
 \rho\tilde{\sigma}=|\delta|\leq\zeta^{-1} \approx 10\, .
\]
At the same time, since $\sinh\pi\approx10$ we will have 
\[
 \rho\tilde{\sigma}\geq 1/\zeta \sqrt{2}\approx 7 \, .
\]
We see that $\rho\tilde{\sigma}$ is not oscillating very much.
Observe also that for moderately large $\nu\approx1$ the angle $\eta$ may 
show considerable oscillations.

\item{\em Midways between cyclotron resonances}, $\omega=n+1/2$. 
Now $\sin\pi(\omega+i\nu)=\pm \cosh\pi\nu$, with still 
$+$ for $n$ even and $-$ for $n$ odd. In this case, the two terms in the 
denominator are both real, which means that for moderate $\nu$ we can get 
a cancellation between the terms, which will make $\rho\tilde{\sigma}$ large, 
and show large oscillations. Since we have neglected terms of order 
$1/\kappa$ we will expect these oscillations to be limited by 
$\rho\tilde{\sigma}<\kappa$. Observe also that in this case we have 
$\eta=-\pi/2$ not oscillating.
\end{enumerate}
We remark that in both cases we have $\rho\tilde{\sigma}\gg1$.
This means that as long as the damping factor $e^{-\pi\nu}$ is larger than 
$1/\beta$ we can neglect the $1/\beta$ terms, and simplify the Fourier 
coefficients further. For $\omega\approx1$ this is the same condition as the 
range of validity of the stationary phase approximation. We obtain
\begin{eqnarray*}
 A_0&=&2\rho\tilde{\sigma}e^{-\pi\nu}\cos2\kappa \sin(\pi\omega+\eta)
 \, , \\
 \tilde{A}_0&=&2\rho\tilde{\sigma}e^{-\pi\nu}\cos2\kappa \sin\pi\omega
 \, , \\
 \tilde{A}_1 &=& -\beta^{-1}\sin\eta-i\rho\tilde{\sigma}\, .
\end{eqnarray*}
We see that all the coefficients are of the same magnitude, except
that $\Re\tilde{A}_1 \ll \Im\tilde{A}$. Looking back to the expression
(\ref{aecurrent}) for the acoustoelectric current, we see that only one
term is relevant, and we get
$${\bf j}_0^{\text{ae}}
  = - \frac{j_0}{\nu^2+1} \beta\rho\tilde{\sigma}
    \left\{\begin{array}{c}\nu\\1\end{array}\right\}\Im\tilde{A}_1.
$$
We can now use equation (\ref{ren}) to calculate the total acoustoelectric 
current. The factor $\gamma$, being the ratio of the Fermi energy to 
the cyclotron energy in the effective magnetic field (times the factor
$1+\nu^2$ which is not too far from 1) will be large. We can then 
approximate ${\bf j}^{\text{ae}}\approx M^{-1}{\bf j}^{\text{ae}}_0$. We 
have 
\[
 \hat{M}^{-1}=\frac{1}{\gamma}
  \left(\begin{array}{cc} 1 &-\nu\\\nu&1\end{array}\right),
\]
which gives
\begin{equation}\label{ae2}
{\bf j}^{\text{ae}}   
 = - j_0 \frac{\beta\rho\tilde{\sigma}}{\gamma}
    \left\{\begin{array}{c}0\\1\end{array}\right\}\Im\tilde{A}_1.
\end{equation}
We see that all the current is in the $y$-direction.

Taking into account all terms we find the ``Hall angle'' given by 
\begin{eqnarray}\label{Hall}
 \tan\theta_H&&=\frac{j_y}{j_x}=-\frac{A_0+\beta|\delta|\Im\tilde{A}_1}
  {|\delta|\tilde{A}_0+\beta|\delta|\Re\tilde{A}_1}\nonumber\\
  &&\approx \frac{\beta|\delta|}{2|\delta|e^{-\pi\nu}\cos2\kappa\sin\pi\omega
  -\sin\eta}.
\end{eqnarray}
The two terms in the denominator are of the same magnitude.

It is customary to relate the acoustoelectric current to the absorption   
of acoustic energy per area.    
This is given by the expression 

\[
 P=\langle\dot{H}f_1\rangle\rightarrow
  \frac{1}{2}\langle\Re\{\dot{H}^*f_1\}\rangle,
\]
where $\langle\cdots\rangle$ denotes average over the period of the 
acoustic wave, and the replacement indicated by the arrow is because of 
the use of complex notation. 

\[
 H=\frac{({\bf P} + \frac{e}{c}{\bf A})^2}{2m} -e \Psi
\]
is the Hamiltonian, and the vector potential is given by $A_x=A_z+0$, 
$A_y=-\delta (c/v_s)\Psi$. This gives rise to the CS magnetic 
field, and the soleniodal part ($y$-component) of the CS electric field. 
We then get 
\[
 P= (g e^2/2)|\Psi|^2\Omega(A_0-\beta\rho\tilde{\sigma}
 \,  \Im\tilde{A}_1),
\]
 where $g=m/2\pi\hbar^2$ is the density of states. 
Comparing with (\ref{ae2}) in the limit $\nu>1$ and within the range
of applicability of the stationary phase approximation we then have
the relation between the acoustoelectric current, acoustic
attenuation, $\Gamma$ and sound intensity, $I$,
\[
 {\bf j}^{\text {ae}}= 
  \left\{\begin{array}{c}0\\\mu_{yx}^t \end{array}\right\}  
   \frac{\Gamma I}{v_s}\, .
\]
 Here we have introduced the so-called traveling-wave
 mobility~\cite{Fritzsche1984,Galperin1991} defined 
 as $\mu^t_{xy} = j^{\text {ae}}_y v_s/\Gamma I$. In the lowest approximation
  $\mu_{yx}^t=1/\rho en_e$ 
 and $n_e=g\epsilon_F$ is the composite fermion 
concentration. Note however that the Weinreich relation 
is only approximate as written here. It is valid only as long as we 
consider all the acoustoelectric current to be in the $y$-direction. 
By measuring the Hall angle one can determine how large a proportion 
of the current is turned by the magnetic field, and then reconstruct 
the Weinreich relation to take this into account.

It is instructive to compare the present expression for the 
Weinreich relation\cite{Weinreich1957} with the one that would be
expected in a normal  
metal, which can be expressed through DC
electron conductivity 
\begin{equation}\label{wcl}
  {\bf j}_{\text {nm}}^{\text {ae}}= 
  \left\{\begin{array}{c}\mu_{xx}^{\text{dc}}\\\mu_{yx}^{\text{dc}}\end{array}\right\}  
   \frac{\Gamma I}{v_s}.
\end{equation}
The mobility $\hat{\mu}^{\text{dc}}$ can be found from the electron
conductivity which, in 
turn,  can be expressed in terms of 
the composite fermion  
DC conductivity (\ref{DCcond}) (see, e.g.,
Ref.~\onlinecite{Simon1998}). As a result, 
\begin{equation}
\hat{ \mu}^{\text{dc}}=\frac{1}{2}\frac{1}{\rho en_e}\left(\begin{array}{cc} 
   \nu/\nu_f^* &-1/\nu_f \\ 1/\nu_f & \nu/\nu_f^*\end{array}\right).
\end{equation}
Here $\nu_f$ is the electron filling factor, which will be close to $1/2$, 
and $\nu_f^*$ is the composite fermion filling factor (The filling 
factor in the effective field $B_0^*$) which is much greater than 1. 
Looking at the $y$-components, we see that the two predictions 
agree at exactly $\nu_f=1/2$. Since $\Gamma$ is some function of 
the magnetic field, it is natural to focus on the traveling-wave
mobility $\mu_{yx}^t \propto j_y^{\text{ae}}/\Gamma$.
The normal metal expression  would predict that this quantity should
increase linearly with increasing magnetic field. Our 
result gives instead a constant value in the region around $\nu_f=1/2$.
In other words, the  mobility $\mu_{yx}^t$ is {\em quantized} close to
$\nu_f=1/2$. 
We can do a similar comparison for the Hall angle. According to (\ref{Hall})
this will be close to $\pi/2$, and we therefore expand around this point.
The normal metal prediction would be 
\[
 |\theta_H-\pi/2|=\sigma_{xx}^e/\sigma_{xy}^e=\nu\nu_f/\nu_f^*=(\omega_c^e \tau)^{-1}   \, , 
\]
which decreases monotonically with increasing magnetic field. Our result 
gives
\[
 |\theta_H-\pi/2|=(\beta|\delta|)^{-1}
\left( 2|\delta|e^{-\pi\nu}\cos2\kappa\sin\pi\omega 
  -\sin\eta \right)\, ,
\]
which will show both geometric oscillations and cyclotron resonance. 
In the limit $\nu_f\to 1/2$ we have $\omega, \nu \gg 1$, which gives
$|\delta|=1/\zeta$ and $\sin\eta=-1$. The Hall angle simplifies 
then to 
\[
 |\theta_H-\pi/2|=(\beta|\delta|)^{-1}=\hbar q/2p_F\, .
\]
This limit is outside the range of applicability of our 
approximation. The result is, however, correct, as can be 
checked by a direct calculation for $\nu_f=1/2$.

By measuring the acoustoelectric current one can use
Eq.~(\ref{aecurrent}) to determine the amplitude $\Phi_0$ of the
effective potential acting upon the composite fermions. As a result,
using the theory\cite{Simon1996} which relates the effective potential
to the coupling constant, one can determine the coupling constant $C$
between the piezoelectric field and the composite fermions. Further,
we note that whereas the non-CS part of the acoustoelectric current
can be expressed in terms of the complex, longitudinal conductivity
for electrons though the Weinreich relation, this is not possible for
the CS part. This means that more information on the system is
available if one can measure the acoustoelectric current compared to
the situation where one only measures attenuation and velocity
shifts. For example, if one measures attenuation and velocity shift,
one can determine the complex longitudinal electronic conductivity,
$\sigma_{xx}^e$, but in general it is impossible to extract the
composite fermion conductivities from this. Measurement of the
acoustoelectric current will give one additional relation which
enables one to determine one more parameter in the composite fermion
theory.

\section{Conclusions}  
  
We have shown that in the cases where the acoustic wave is sensitive
to the properties of the electron gas, the AC magnetic Chern-Simons
field created by the induced density fluctuations in the electron gas
creates forces that are of strength comparable to the direct
perturbation, and that the same is true for the $x$-component of the
electric Chern-Simons field, The $y$-component of this field is
smaller but plays an important role because it acts along the
composite fermion trajectory at the points of strong
interaction. At the second order, necessary for
the calculation of the acoustoelectric current, the Chern-Simons field
will be relevant, and we expect a violation of the Weinreich relation.
This violation can be detected in measurements of the Hall angle.
This gives additional information on the composite fermion system 
not attainable by linear response measurements. Further, the 
acoustoelectric current gives a direct measure of the strength of 
the piezoelectric field as seen by the composite fermions, and 
may therefore be used to extract the value of the coupling constant
between the acoustic wave and the composite fermions. 
  
\appendix  
\section{Asymptotic expansions}\label{asymptotic}  
The stationary phase approximation is used to extract the leading term
in the asymptotic expansions both for the conductivity and the
acoustoelectric current. In general, if we have a function

\begin{equation} F(\kappa)=\int_a^bf(x)e^{i\kappa g(x)}\, dx \, ,
\end{equation}  
the leading terms in the asymptotic expansion will be   
\begin{eqnarray*}  
 F(\kappa)&=&F^{(1)}(\kappa)+F^{(2)}(\kappa);  
\\ 
 F^{(1)}(\kappa)&=&\sqrt{\frac{2\pi}{\kappa}}\sum_{j=1}^n\frac{e^{\pm
i\pi/4}}   
  {\sqrt{|g''(x_j)|}}f(x_j)e^{i\kappa g(x_j)}\, ,  
\end{eqnarray*} 
where the sum is over all the stationary points of the exponent 
($g'(x_j)=0$),    
and the sign is $\pm$ according as $g''(x_j)\gtrless 0$.  
$$  
 F^{(2)}(\kappa)=\frac{1}{i\kappa}\left[\frac{f(b)}{g'(b)}e^{i\kappa g(b)}  
   -\frac{f(a)}{g'(a)}e^{i\kappa g(a)}\right]  
$$  
is the contributions from the end points (see, e.g., Ref.~\onlinecite{Mandel}).
 For example to find $B_1$ we have to evaluate
 \begin{equation}  
 I=\int_0^{2\pi} d\phi e^{-i\phi} f_1(\phi)\, .  
\end{equation}  
Inserting the expression (\ref{sle}) for $f_1$ we get, among other terms   
the integral
\begin{eqnarray}  
 I&=&M\int_0^{2\pi}d\phi\, \cos\phi\int_{\phi}^{\phi+2\pi}d\phi'\, 
   e^{\nu(\phi'-\phi) 
  +i\kappa[g(\phi)-g(\phi')]}\, , \nonumber \\  
 M&=&i\omega\, e^{-2\pi(\nu-i\omega)},\quad   
 g(\phi)=\sin\phi-\phi/\beta \, .  
\end{eqnarray}  
Performing first the integral over $\phi'$ we get  
 \begin{eqnarray}  
 I'&=&\int_{\phi}^{\phi+2\pi}d\phi\, 
 e^{\nu\phi'+i\kappa[\sin\phi'-   
  \phi'/\beta]}\nonumber \\  
 &\sim& \sqrt{\frac{2\pi}{\kappa}}\sum_je^{\nu\phi_j'+i\kappa g(\phi_j')  
  -\frac{i\pi}{4}\mbox{\small sign}(\sin\phi')}\nonumber \\  
 &&-\frac{i}{\kappa}\frac{1}{\cos(\phi)-1/\beta}\,  
  e^{\nu\phi+i\kappa g(\phi)}\left[e^{2\pi(\nu-i\omega)}-1\right] \, .  
\end{eqnarray}  
Here $\phi_j'$ are the stationary points of $g(\phi')$,
$\cos\phi_j'=1/\beta$,   
which are in the range $\phi\leq\phi_j'\leq\phi+2\pi$. The integral over   
$\phi$ can then be evaluated as   
 \wide
\begin{eqnarray}  
 I&=&M\int_0^{2\pi}d\phi\cos\phi e^{-\nu\phi-i\kappa g(\phi)}  
   \sum_je^{\nu\phi_j'+i\kappa g(\phi_j')  
   -\frac{i\pi}{4}\mbox{\small 
   sign}(\sin\phi')}+\frac{1}{\beta}\int_0^{2\pi}d\phi   
   \frac{\cos\phi}{\cos\phi-1/\beta}\nonumber \\  
  &=&\frac{2\pi M}{\kappa\beta}\sum_{i,j}e^{\nu(\phi_j'-\phi_i)  
   +i\kappa(g(\phi_j')-g(\phi_i))-\frac{i\pi}{4}  
   (\mbox{sign}(\sin\phi_j')-\mbox{\small sign}(\sin\phi_i))}  
   +\frac{1}{\beta}\int_0^{2\pi}d\phi  
   \frac{\cos\phi}{\cos\phi-1/\beta}\, ,\nonumber  
\end{eqnarray}  
\narrow  
where the critical points $\phi_i$ are the same as before.  
In the sum over the critical points we can then set $\phi_i$, $\phi_j'$   
equal to $\pm\pi/2+2\pi n$ in the exponent since $\beta\gg 1$.  
Also, critical points that lie on the boundary of the domain of   
integration is given half their normal value.   
In the second integration, the contributions from the end points have   
been neglected, as they will be of higher order.   
The sum then   
 $$  
 \sum_{i,j}(\cdots)=1+e^{2\pi(\nu-i\omega)} 
  +2e^{\pi(\nu-i\omega)}\sin2\kappa\,.  
$$  
  The last integral must be expanded in powers of $1/\beta$, yielding   
 $$  
 \int_0^{2\pi}d\phi\frac{\cos\phi}{\cos\phi-1/\beta}  
  = \int_0^{2\pi}d\phi\left(1+\frac{1}{\beta\cos\phi} + \ldots\right)  
  = 2\pi \, ,  
$$  
since   
$  
 P\int_0^{2\pi}\! d\phi/\cos\phi =0  
$,  
where $P$ is the principal value. Combining these results we get  
one contribution to $B_1$, the rest of   
the $B_n$ being calculated in a similar way.

\widetext  
\end{document}